\documentclass[10pt,journal,compsoc]{IEEEtran}

\usepackage{algorithmic}
\usepackage{array}
\usepackage[nocompress]{cite}
\usepackage{color}
\usepackage{listings}
\usepackage[caption=false,font=normalsize,labelfont=sf,textfont=sf]{subfig}
\usepackage{stfloats}
\usepackage{url}
\usepackage{multirow}

\definecolor{lightgray}{rgb}{0.95,0.95,0.95}

\usepackage{listings}
\usepackage[pdftex]{graphicx}
\DeclareGraphicsExtensions{.pdf,.jpeg,.png}

\usepackage{amsmath}
\begin{document}

\title{Reproducible Workflow on a Public Cloud for Computational Fluid Dynamics}
\author{Olivier Mesnard, Lorena A. Barba
\IEEEcompsocitemizethanks{\IEEEcompsocthanksitem Mechanical and Aerospace Engineering,
the George Washington University, Washington, DC 20052.\protect\\
E-mail: mesnardo@gwu.edu
\IEEEcompsocthanksitem Email: labarba@gwu.edu}
}

\IEEEtitleabstractindextext{%
\begin{abstract}
In a new effort to make our research transparent and reproducible by others, we developed a workflow to run and share computational studies on the public cloud Microsoft Azure.
It uses Docker containers to create an image of the application software stack.
We also adopt several tools that facilitate creating and managing virtual machines on compute nodes and submitting jobs to these nodes.
The configuration files for these tools are part of an expanded ``reproducibility package'' that includes workflow definitions for cloud computing, in addition to input files and instructions.
This facilitates re-creating the cloud environment to re-run the computations under the same conditions.
Although cloud providers have improved their offerings, many researchers using high-performance computing (HPC) are still skeptical about cloud computing.
Thus, we ran benchmarks for tightly coupled applications to confirm that the latest HPC nodes of Microsoft Azure are indeed a viable alternative to traditional on-site HPC clusters.
We also show that cloud offerings are now adequate to complete computational fluid dynamics studies with in-house research software that uses parallel computing with GPUs.
Finally, we share with the community what we have learned from nearly two years of using Azure cloud to enhance transparency and reproducibility in our computational simulations.
\end{abstract}
}

\maketitle

\IEEEraisesectionheading{\section{Introduction}\label{sec:introduction}}

\IEEEPARstart{R}{eproducible} research and replication studies are essential components for the progress of evidence-based science, even more now when nearly all fields advance via computation.
We use computer simulations and data models to create new knowledge,
but how do we provide evidence that this new knowledge is justified? 
Traditional journal publications exclude software and data products from the peer-review process, yet reliance on ever more complex computational artifacts and methods is the norm.
Lacking standards for documenting, reporting and reviewing the computational facets of research, it becomes difficult to verify and corroborate the findings presented in journals and conferences \cite{donoho_et_al_2009}.

The literature is cluttered with confused and sometimes contradictory definitions for reproducible research, reproducibility, replicability, repetition, etc.\cite{barba_2018}.
It is thus worth clarifying how we use these terms.
``Reproducible research'' was used by geophysics professor Jon Claerbout in the 1990s to mean computational studies that are published with sufficient transparency so other scientists can re-create the results. His research group at Stanford created a reproducible-research environment\cite{schwab_et_al_2000} whose goal was complete documentation of scientific computations, in such a way that a reader could reproduce all the results and figures in a paper using the author-provided computer programs and raw data.
This requires open data and open source software and, for this reason, the reproducibility movement is closely linked with the open science movement.
The term ``replication'' has been adopted to refer to an independent study generating new data which, when analyzed, lead to the same findings \cite{peng_2011}.
We follow this convention, adopted also recently in reports from the National Academies of Sciences, Engineering, and Medicine \cite{nasa_oss_2018,nasem_2019}.

Many efforts to develop cyberinfrastructure that supports reproducible research have been launched in past years. 
They address concerns like automatic capture of changes to software (version control systems), persistent data archival, global registration of data identifiers, workflow management, and more. 
But capturing the whole computational environment used in a research project remains one of the most difficult problems. 
Computational researchers often use a multi-layer stack of software applications that can be laborious to build from scratch. 
Container technology like Docker is a recent addition to the reproducibility toolbox \cite{boettiger_2015}. 
In this work, we develop and assess a workflow for reproducible research on the public cloud provider Microsoft Azure, adopting Docker containers and several other tools to automate and fully document scientific computations.
The workflow follows the idea of using container technology to enhance transparency and reproducibility, and to reduce the burden of installing our application software stack.
We go one step further by moving the computational work to Azure cloud resources using tools that programmatically capture the virtual environment configuration.
The configuration files become part of expanded ``reproducibility packages'' for the study.
This work does not target portability of our software and workflow to diverse high-performance computing (HPC) environments. 
Our goal is, rather, to publish our computational research with deep transparency. 
The workflow we developed and document here is expressly designed for Azure  cloud offerings. 
An interested reader could thus re-run our simulations under exactly the same conditions as we used, provided they are willing and able to create an Azure account.

Universities and national laboratories spend millions of dollars to deploy and maintain on-site HPC clusters.
At the George Washington University, we have access to a cluster called Colonial One.
The cluster is now $6$ years old and approaching its end-of-life.
On average, its computational resources are idle $9\%$ of the time and unavailable to users for roughly 5 days a year (due to maintenance).
Administrators of the cluster have been recently considering integrating cloud-computing platforms in the research-computing portfolio. 
Over the past decade, cloud-computing platforms have rapidly evolved, now offering solutions for scientific applications.
From a user's point of view, a cloud platform offers great flexibility (hardware and virtual machines) with instantaneous availability of infinite (in appearance) computational resources.
No more waiting time in job-submission queues!
This promises to greatly facilitate code development, debugging, and testing. Resources allocated on the cloud are released as soon as the job is done, avoiding paying for idle time.
On a public-cloud platform---such as Microsoft Azure, Google Cloud, Amazon AWS---with a ``pay-as-you-go'' type of subscription, a user directly sees how much it costs to run a scientific application, while such information usually remains obscure to the end-user on university-managed clusters. 
Cost models that make sense for researchers, labs, and universities are still unclear. 
Yet, this information is key when making a decision to adopt a cloud workflow for research. 

Until recently, public cloud offerings were inadequate to the needs of computational scientists using HPC, mainly due to performance overhead of virtualization or lack of support for fast networking \cite{freniere_et_2016} and hardware accelerators.
Freniere and co-authors (2016) used the OSU micro-benchmark suite to report performance degradation in networking with Amazon AWS, compared to their on-site local HPC cluster.
We take the same approach with Microsoft Azure to verify that their improved services are now suitable for HPC applications.
We also share what we have learned from nearly two years of using cloud computing for our computational fluid dynamics (CFD) simulations, with the hope that it will shed some light on the costs and benefits, and help others considering using cloud computing for their research.

\section{Reproducible cloud-based workflow}\label{sec:workflow}

Scientific publications reporting computational results often lack sufficient details to reproduce the researcher's computational environment; e.g., they may miss to mention external libraries used along with the main computational code.
We have learned the hard way how different versions of the same external library can alter the numerical results and even the scientific findings of a computational study\cite{mesnard_barba_2017}.
This section presents an overview and mini-tutorial of the workflow we developed to aim for the highest level of reproducibility of our computational research, with the best available technology solutions. 
In the process of creating this reproducible workflow, we also evaluated the suitability of public cloud offerings by Microsoft Azure for our research computing needs. 
The tools we adopted for computing on cloud resources are specific for this provider. 

\subsection{Use of Container Technology}

\begin{figure*}
    \begin{center}
        \colorbox{lightgray}{
        \begin{minipage}[c]{14cm}
        \bigskip
        \small \sffamily{
        \subsection*{Docker Terminologies:}
{\sf \footnotesize
\textbf{Docker}--- An open source OS-level virtualization software to create and run multiple independent, isolated, and portable containers on the same host Operating System.\\
\vspace{0.1cm}\\
\textbf{Docker Image}--- Union of layered filesystems stacked on top of each other. Each layer defines a set of differences from the previous layer. A user composes (builds) a Docker image using a \texttt{Dockerfile}, usually starting from a base image (such as \texttt{ubuntu:16.04}).\\
\vspace{0.1cm}\\
\textbf{Docker Container}--- A standardized unit created from a Docker image to deploy an application or a runtime environment. A Docker container can be seen as an instance of a Docker image that includes an additional writable layer at the top of the layered stack. When a container is deleted, so is the writable layer, while the image remains unchanged.\\
\vspace{0.1cm}\\
\textbf{Dockerfile}--- An ASCII file including the sequence of instructions to create a Docker image for the computational runtime environment. A Dockerfile contains Docker keywords such as \texttt{FROM}, \texttt{RUN}, or \texttt{COPY}. Each instructions in the \texttt{Dockerfile} creates a layer in the Docker image.\\
\vspace{0.1cm}\\
\textbf{DockerHub}--- The official registry of Docker Inc.; a cloud-based registry service to store, share, and retrieve (public or private) Docker images.\\
}}
        \vspace{0.2cm}
        \end{minipage}}
    \end{center}
\end{figure*}

To overcome the so-called ``dependency hell'' and facilitate reproducibility and portability, we use the container technology provided by the open-source project Docker.
A container represents an isolated user space where application programs run directly on the operating system's kernel of the host (with limited access to its resources).
In contrast with virtual machines, containers do not include a full operating system, making them lighter and faster.
Containers allow re-creating the same runtime environment of an application (including all its dependencies) from one machine to another. 
They empower researchers to share pre-built images of their software stack, ameliorating one of the biggest pain points for reproducible research: building the software on a different machine. 
The majority of enterprise developers today are familiar with Docker container technology (first released six years ago). 
But in academic settings, many are still unaware of its value. 
We present this section as an overview for research software developers unfamiliar with Docker, but comfortable with scripting, version control, and distributed collaboration.

A container is an instance of an image.
The developer builds this image on a local machine and pushes it to a public registry to share it with other users.
Users pull the image from the public registry,---in our case, DockerHub---and create containers out of it.
To create a Docker image, the developer writes a \texttt{Dockerfile}: an ASCII file containing instructions that tell Docker what to do.
For example, we start building a new image from a base image using the keyword \texttt{FROM}.
We then write the different shell instructions (prefixed with the keyword \texttt{RUN}) to build a multi-layered runtime environment of the application that includes all its dependencies.
Once we have built and tested the image on the local machine, we push it to a repository on the DockerHub registry: a place to store and retrieve Docker images.
Listing \ref{lst:docker_build_push} provides the command lines to build (\texttt{docker build}) and push (\texttt{docker push}) an image of our CFD software (\texttt{barbagroup/petibm:0.4-GPU-IntelMPI-ubuntu}) that we used to obtain some of the results presented in the next section.
Here, \texttt{CLOUDREPRO} is an environment variable set to the local path of the GitHub repository for this paper, \texttt{cloud-repro},\footnote{\url{https://github.com/barbagroup/cloud-repro}} cloned on the user's machine.

\begin{lstlisting}[label=lst:docker_build_push,caption={Build and push a Docker image.}]
$ cd $CLOUDREPRO/docker/petibm
$ docker build --tag=barbagroup/petibm:0.4-GPU-IntelMPI-ubuntu --file=Dockerfile .
$ docker push barbagroup/petibm:0.4-GPU-IntelMPI-ubuntu
\end{lstlisting}

\begin{figure*}[t]
    \centering
    \includegraphics[width=16cm]{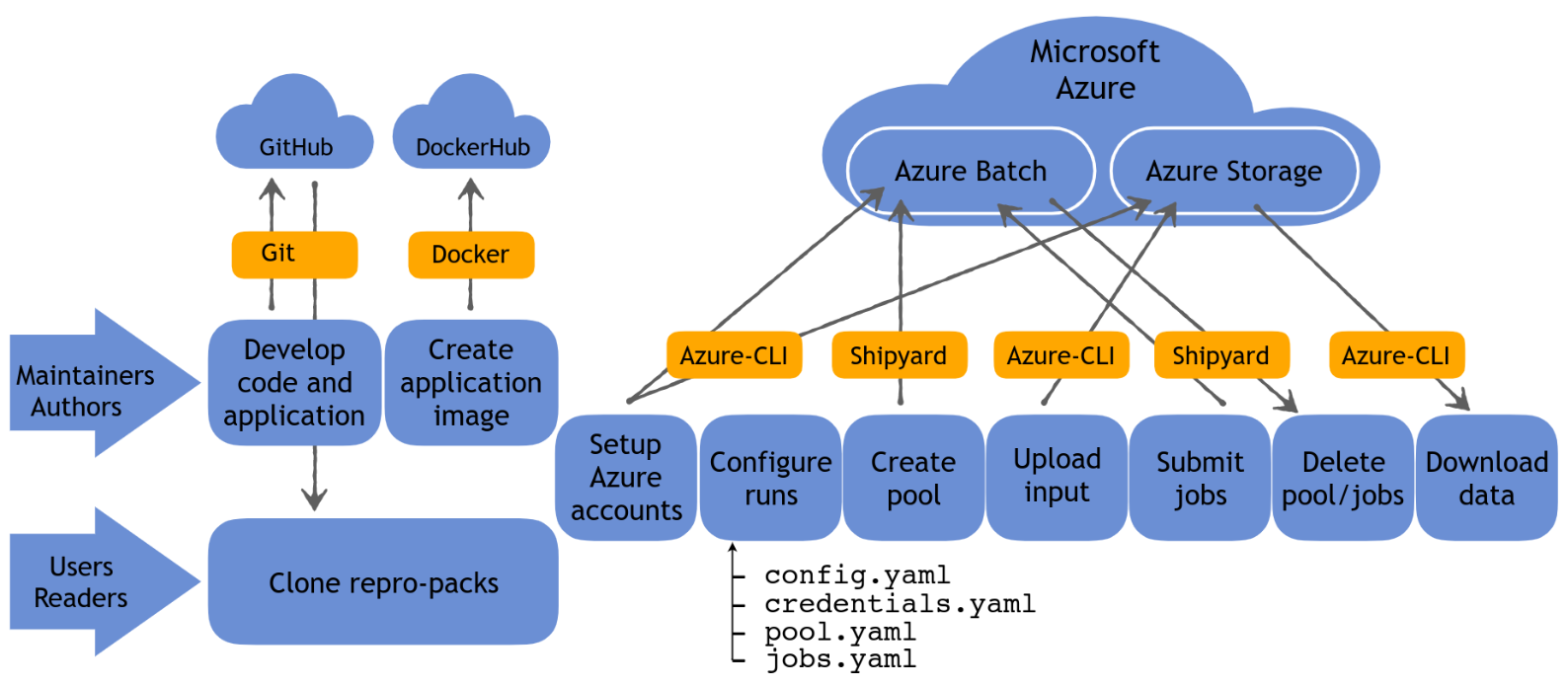}
    \caption{Reproducible workflow on the public cloud provider Microsoft Azure. Our CFD software is version-controlled with Git and GitHub. We push to DockerHub a Docker image of our CFD application with all its dependencies. Azure CLI is used to configure accounts on Microsoft Azure and to upload/download data to/from an Azure Storage account. With Batch Shipyard, we create a pool on Azure Batch and run container-based simulations using our Docker image. Figure available under CC BY 4.0 license (https://doi.org/10.6084/m9.figshare.9636722.v1).}
    \label{fig:cloud_workflow}
\end{figure*}

\noindent A reader interested in reproducing the computational results can now pull the application image from DockerHub, and create a Docker container to run the CFD application software in a faithfully reproduced local environment.
Our objective is to create and run containers on the public cloud provider Microsoft Azure, and make it possible for the readers to also run these containers to reproduce our results.
Figure \ref{fig:cloud_workflow} shows a graphical representation of the workflow we developed using Docker and various tools for running CFD simulations on Microsoft Azure.
The next section explains these tools.

\subsection{Use of Public Cloud Resources}

To run computational jobs on Microsoft Azure, we use several tools that facilitate creating and managing virtual machines on compute nodes, and submitting jobs to those nodes.
We use a service called Azure Batch that leverages Microsoft Azure at no extra cost, relieving the user from manually creating, configuring, and managing an HPC-capable cluster of cloud nodes, including virtual machines, virtual networks, job and task scheduling infrastructure.
Azure Batch works with both embarrassingly parallel workloads and tightly coupled MPI jobs (the latter being the case of our CFD software).
To use Azure Batch, we first need to configure a workspace on Microsoft Azure.
This can be done either via the Azure Portal in a web browser or from a local terminal using the open-source tool Azure CLI.\footnote{Azure CLI (version 2.0.57): \url{https://github.com/Azure/azure-cli}}
We prefer to use the command-line solution (program \texttt{az}), as it allows us to keep track of the steps taken to configure the cloud workspace (see Listing \ref{lst:az_configure}).
First, we set the Azure subscription we want to use (let's call it \texttt{reprosubscription}).
Next, we create a resource group (\texttt{reprorg}) located in this case in the East US region, which will contain all the Azure resources.
We create an Azure Storage account (\texttt{reprostorage}) in the resource group, as well as an Azure Batch account (\texttt{reprobatch}) associated to the storage account.
Finally, we create a fileshare (in this case of size $100$ GB) in the storage.

\begin{lstlisting}[label=lst:az_configure,caption={Configure the workspace on Microsoft Azure.}]
$ az account set --subscription reprosubscription
$ az group create --name reprorg --location eastus
$ az storage account create --name reprostorage --resource-group reprorg --sku Standard_LRS --location eastus
$ az batch account create --name reprobatch --resource-group reprorg --location eastus --storage-account reprostorage
$ az storage share create --name fileshare --account-name reprostorage --account-key storagekey --quota 100
\end{lstlisting}

To create computational nodes and submit container-based jobs to Azure Batch, we use the open-source command-line utility Batch Shipyard.\footnote{Batch Shipyard (version 3.6.1): https://github.com/Azure/batch-shipyard}
Batch Shipyard is entirely driven by configuration files: the utility parses user-written YAML files to automatically create pools of compute nodes on Azure Batch and to submit jobs to those pools.
Typically, we need to provide four configuration files:

\begin{itemize}
    \item \texttt{config.yaml} contains information about the Azure Storage account and Docker images to use.
    \item \texttt{credentials.yaml} stores the necessary credentials to use the different Microsoft Azure service platforms (e.g., Azure Batch and Azure Storage).
    \item \texttt{pool.yaml} is where the user configures the pool of virtual machines to create.
    \item \texttt{jobs.yaml} details the configuration of the jobs to submit to the pool.
\end{itemize}

Once the configuration files are written, we invoke Batch Shipyard (program \texttt{shipyard}) on our local machine.
The folder \texttt{examples/snake2d2k35/config\_shipyard} in the repository accompanying this paper contains an example YAML files to create a pool of two \texttt{NC24r} compute nodes (featuring K80 GPUs and using InfiniBand network).
Listing \ref{lst:shipyard_run} shows the commands to run in your local terminal to create a pool of compute nodes on Azure Batch and submit jobs to it.
The Docker image of our CFD application is pulled from the registry to the virtual machines during the pool creation (\texttt{shipyard pool add}).
We then upload the input files to the compute nodes (\texttt{shipyard data ingress}) and submit jobs to the pool (\texttt{shipyard jobs add}). The tasks for a job will start automatically upon submission.

\begin{lstlisting}[label=lst:shipyard_run,caption={Create a pool and submit jobs to it.}]
$ cd $CLOUDREPRO/examples/snake2d2k35
$ az storage directory create --name snake2d2k35 --share-name fileshare --account-name reprostorage
$ export SHIPYARD_CONFIGDIR=config_shipyard
$ shipyard pool add
$ shipyard data ingress
$ shipyard jobs add
\end{lstlisting}

Once the simulations are done (i.e., the job tasks are complete), we delete the jobs and the pool (Listing \ref{lst:shipyard_del}).
The output of the computation is now stored in the fileshare in our Azure Storage account.
We can download the data to our local machine to perform additional post-processing steps (such as flow visualizations).

\begin{lstlisting}[label=lst:shipyard_del,caption={Delete the pool and jobs, and download to output to a local machine.}]
$ shipyard pool del
$ shipyard jobs del
$ mkdir output
$ az storage file download-batch --source fileshare/snake2d2k25 --destination output --account-name reprostorage
\end{lstlisting}

Reproducible research requires authors to make their code and data available.
Thus, the Dockerfile and YAML configuration files should be made part of an extended reproducibility package that includes workflow instructions for cloud computing, in addition to other input files.
Such a reproducibility package facilitates re-creating the cloud environment to run the simulations under the same conditions.
The reproducibility packages of the examples showcased in the next section are available in the GitHub repository \texttt{cloud-repro}, which includes instructions on how to reproduce the results.

\section{Results}\label{sec:results}

The top concerns of researchers considering cloud computing are performance and cost. 
Until just a few years ago, the products offered by cloud providers were unsuitable to the needs of computational scientists using HPC, due to performance overhead of virtualization or lack of support for fast networking \cite{freniere_et_2016}.
Azure only introduced nodes with GPU devices during late 2016 and Infiniband support for Linux virtual machines on the NC-series in 2017.
Our first objective was to assess performance on cloud nodes for the type of computations in our research workflows with  tightly coupled parallel applications. 
We present results from benchmarks and test-cases showing that we are able to obtain similar performance in terms of latency and bandwidth using the Azure virtual network, comparing to a traditional university-managed HPC cluster (Colonial One).
Table \ref{tab:hw_specs} lists the hardware specifications of the nodes used on Microsoft Azure and Colonial One.
Our target research application relies on three-dimensional CFD simulations with our in-house research software. We include here a sample of the types of results needed to answer our research question, obtained by running on the public cloud using the reproducible workflow described in Section \ref{sec:workflow}.
The goal is to showcase the potential of cloud computing for CFD, share the lessons we learned in the process, as well as analyze the cost scenarios for full applications.

\begin{table*}[b]
    \renewcommand{\arraystretch}{1.5}
    \caption{Hardware specifications of nodes used on Microsoft Azure and Colonial One. On both platforms, MPI applications take advantage of RDMA (Remote Direct Memory Access) network with FDR InfiniBand and ECC is enabled for GPU computing.}
    \label{tab:hw_specs}
    \centering
    \begin{tabular}{cccccccc}
        Platform & Node & Intel Xeon CPU & \# threads & NVIDIA GPU & RAM (GiB) & SSD Storage (GiB) \\
        \hline
        \multirow{2}{*}{Azure} & \texttt{NC24r} & Dual 12-Core E5-2690v3 (2.60GHz) & 24 & 2 x K80 & 224 & 1440 \\
        & \texttt{H16r} & Dual 8-Core E5-2667v3 (3.20GHz) & 16 & - & 112 & 2000 \\
        \hline
        \multirow{2}{*}{Colonial One} & \texttt{Ivygpu} & Dual 6-Core E5-2620v2 (2.10GHz) & 12 & 2 x K20 & 120 & 93 \\
        & \texttt{Short} & Dual 8-Core E5-2650v2 (2.60GHz) & 16 & - & 120 & 93 \\
        \hline
    \end{tabular}
\end{table*}

\subsection{MPI Communication Benchmarks}\label{subsec:mpi_benchmarks}

\begin{figure}[!h]
    \centering
    \includegraphics[width=8cm]{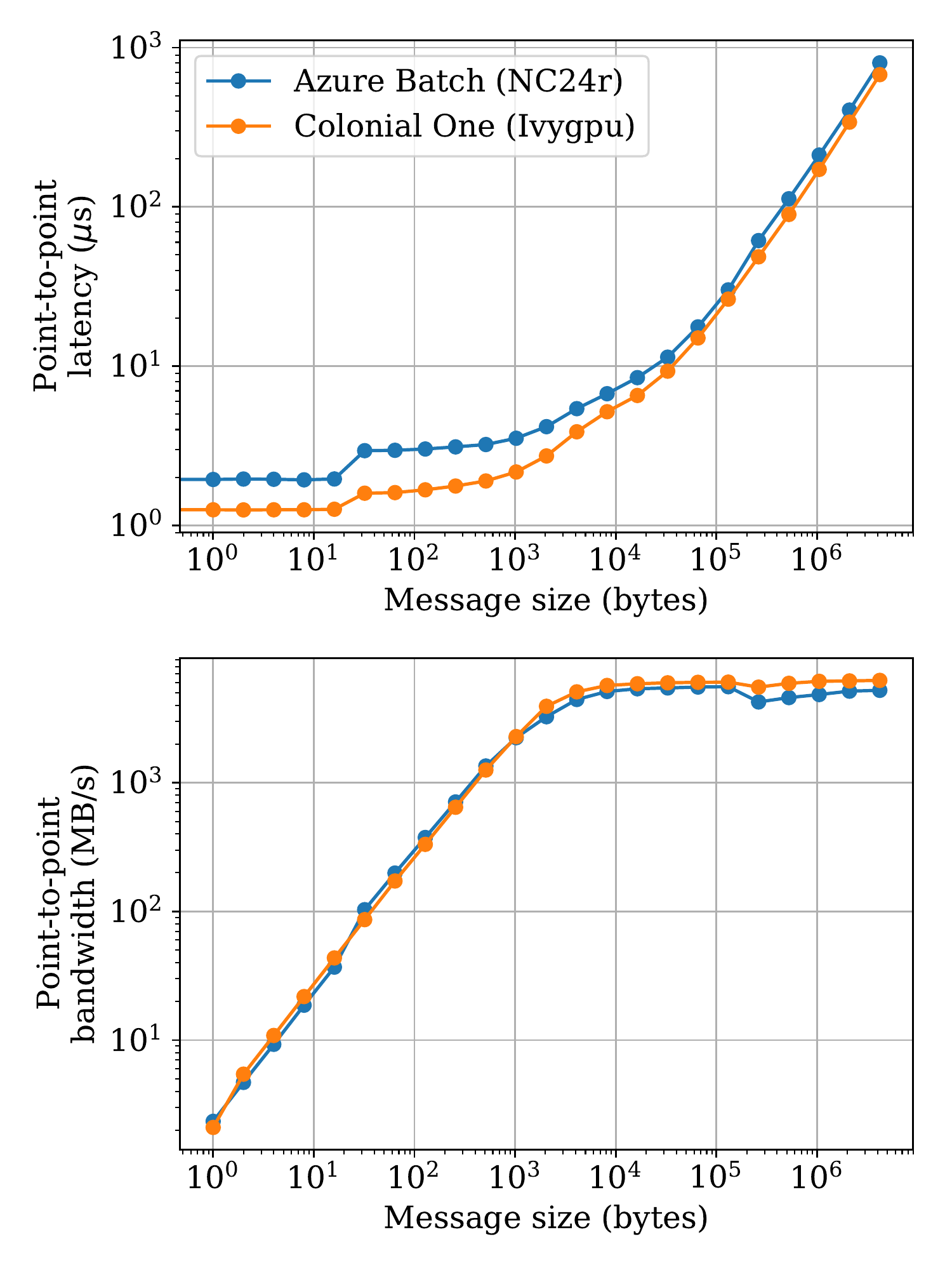}
    \caption{Point-to-point latency (top) and bandwidth (bottom) obtained on Colonial One (\texttt{Ivygpu} nodes) and on Microsoft Azure (\texttt{NC24r} nodes) with the Intel MPI library. Benchmark results are averaged over $5$ repetitions.}
    \label{fig:osu_benchmarks}
\end{figure}

We ran point-to-point MPI benchmarks from the Ohio State University Micro-Benchmarks suite\footnote{OSU Micro-Benchmarks (version 5.6): \url{http://mvapich.cse.ohio-state.edu/benchmarks/}} on Microsoft Azure and Colonial One, to investigate performance in terms the latency and bandwidth.
The latency test is carried out in a ping-pong fashion and measures the time elapsed to get a response; the sender sends a message with a certain data size and waits for the receiver to send back the message with the same data size.
The bandwidth test measures the maximum sustained rate that can be achieved on the network; the sender sends a fixed number of messages to a receiver that replies only after receiving all of them.
The tests ran on \texttt{NC24r} nodes on Azure and \texttt{Ivygpu} nodes on Colonial One, all of them featuring a network interface for RDMA (Remote Direct Memory Access) connectivity to communicate over InfiniBand.
(RDMA allows direct access to a remote's memory without involving the operating system of the host and remote.)
Fig. \ref{fig:osu_benchmarks} reports the mean latencies and bandwidths obtained over $5$ repetitions on both platforms.
For small message sizes, the average latencies on Colonial One and Azure are $1.25$ and $1.95 \mu $s, respectively.
For all message sizes, the latency reported on Azure is slightly higher than the value obtained on Colonial One.
The maximum sustained bandwidth rates for Colonial One and Azure are on average $6.2$ and $5.2$ GB/s, respectively.
For all message sizes, a similar bandwidth rate was achieved on Azure and Colonial One.
Over the last few years, Microsoft Azure has indeed improved its HPC solutions to provide networking capabilities that are comparable or even better than our 6-year-old university-managed cluster.

\subsection{Poisson Benchmarks}\label{subsec:poisson_benchmarks}

\begin{figure}[!h]
    \centering
    \includegraphics[width=8cm]{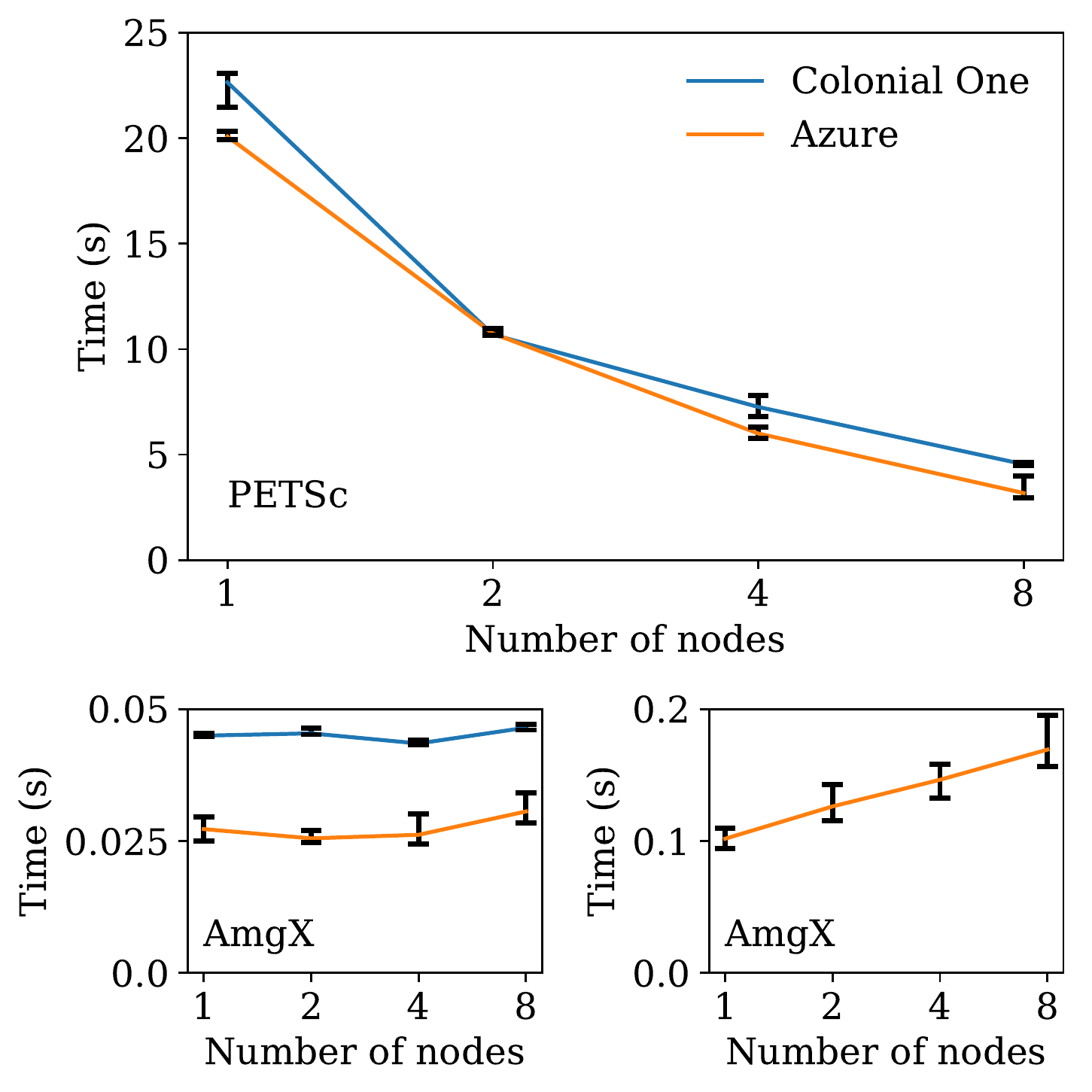}
    \caption{Runtime to solve a Poisson system on Colonial One and Microsoft Azure. Benchmarks were repeated 5 times and we show the mean runtime and the extrema. \textbf{Top}: system solved on a fixed mesh size of 50 million cells (strong scaling) using PETSc on Azure \texttt{H16r} nodes ($16$ processes per node), and Colonial One \texttt{Short} nodes ($16$ processes per node). \textbf{Bottom-left}:  systems solved using AmgX on Azure \texttt{NC24r} nodes ($12$ processes and $2$ GPUs per node), and Colonial One \texttt{Ivygpu} nodes ($12$ processes and $2$ GPUs per node); the base mesh size contains 6.25 million cells and we scale it with the number of nodes (weak scaling). \textbf{Bottom-right:} systems solved with AmgX on Azure \texttt{NC24r} nodes ($24$ processes and $4$ GPUs per node), using a finer base mesh size of 25 million cells. Runtimes obtained with AmgX were normalized by the number of iterations required to reach a absolute tolerance of $10^{-12}$.}
    \label{fig:poisson_benchmarks}
\end{figure}

CFD algorithms often require solving linear systems with iterative methods.
For example, the Navier-Stokes solver implemented in our software requires the solution of a Poisson system at every time step to project the velocity field onto the divergence-free space (to satisfy the incompressibility constraint).
We investigated the time-to-solution for a three-dimensional Poisson system (obtained with a 7-point stencil central-difference scheme) on different nodes of Microsoft Azure and Colonial One.
The solution method was a Conjugate-Gradient (CG) method with a classical algebraic multi-grid (AMG) preconditioner using an exit criterion set to an absolute tolerance of $10^{-12}$.
The iterative solver ran on CPU nodes (\texttt{H16r} instances on Azure and \texttt{Short} nodes on Colonial One) using the CG algorithm from the PETSc library\cite{balay_et_al_2018} and an AMG preconditioner from Hypre BoomerAMG.
Fig. \ref{fig:poisson_benchmarks} (top) reports the mean runtimes (averaged over $5$ repetitions) to iteratively solve the system, on a uniform grid of $50$ million cells ($1000 \times 1000 \times 50$), as we increase the number of nodes in the pool (strong scaling).
Runtimes obtained on Colonial One and Azure are similar.
We also solved the Poisson system with the NVIDIA AmgX library on multiple GPU devices using \texttt{NC24r} instances on Azure and \texttt{Ivygpu} nodes on Colonial One.
The Poisson system for a base mesh of $6.25$ million cells ($500 \times 500 \times 25$) was solved on a single compute node using $12$ MPI processes and $2$ GPU devices; we then doubled the mesh size as we doubled the number of nodes, keeping the same number of MPI processes and GPUs per node (weak scaling).
The number of iterations to reach convergence increases with the size of the system, so we normalize the runtimes by the number of iterations.
Fig. \ref{fig:poisson_benchmarks} (bottom left) shows the normalized mean runtimes ($5$ repetitions) obtained on Azure and Colonial One.
The smaller runtimes on Azure are explained by the fact that the NC-series of Microsoft Azure features NVIDIA K80 GPU devices with a higher compute capability than the K20 GPUs on Colonial One.
The bottom-right panel of the figure reports the normalized mean runtimes obtained on Azure when we load the bandwidth with a larger problem; the base mesh now contains $25$ millions cells ($1000 \times 500 \times 50$).
We observe larger variations in the time-to-solution for the Poisson system on Microsoft Azure with GPU computing; runtimes are more uniform on Colonial One.

\subsection{Flow Around a Flying Snake Cross-Section}\label{subsec:snake}

Our research lab is interested in understanding the aerodynamics of flying animals via CFD simulations.
One of our applications deals with the aerodynamics of a snake species, \textit{Chrysopelea paradisi}, that lives in South-East Asia.
This arboreal reptile has the remarkable capacity to turn its entire body into a wing and glide over several meters\cite{socha_2011}.
The so-called ``flying snake'' jumps from tree branches, undulates in the air, and is able to produce lift by expanding its ribcage to flatten its ventral surface (morphing its normally circular cross-section into a triangular shape).

To study the flow around the flying snake, we developed a CFD software called PetIBM\cite{chuang_et_al_2018}, an open-source toolbox that solves the two- and three-dimensional incompressible Navier-Stokes equations using a projection method (seen as an approximate block-LU decomposition of the fully discretized equations\cite{perot_1993}) and an immersed-boundary method (IBM).
Within this framework, the fluid equations are solved over an extended grid that does not conform to the surface of the body immersed in the computational domain.
To model the presence of the body, the momentum equation is augmented with a forcing term that is activated in the vicinity of the immersed boundary.
This technique allows the use of simple fixed structured Cartesian grids to solve the equations.
PetIBM implements immersed boundary methods (IBMs) that fit into the projection method; in the present study, we use the IBM scheme proposed in \cite{li_et_al_2016}.
PetIBM runs on distributed-memory architectures using the efficient data structures and parallel routines from the PETSc library.
The software also implements the possibility to solve linear systems on multiple GPU devices distributed across the nodes with the NVIDIA AmgX library and our AmgXWrapper\cite{chuang_barba_2017}.
One of the requirements for reproducible computational results is to make the code available under a public license (ideally allowing reuse and modification by others).
In that regard, PetIBM is open source, version-controlled on GitHub,\footnote{PetIBM (version 0.4): \url{https://github.com/barbagroup/PetIBM}} and shared under the permissive (non copy-left) BSD-3 clause license.
We also provide a Docker image of PetIBM on DockerHub and its Dockerfile is available in the GitHub repository of the software.

\subsubsection{2D Flow Around a Snake Cross-Section}

We submitted a job on Azure Batch (with Batch Shipyard) to compute the two-dimensional flow around an anatomically accurate sectional shape of the gliding snake.
The cross-section has a chord-length $c=1$ and forms a $35$-degree angle of attack with the incoming freestream flow.
The Reynolds number, based on the freestream speed, the body chord-length, the kinematic viscosity, is set to $Re=2000$.
The immersed boundary is centered in a $30c \times 30c$ computational domain that contains just over $2.9$ million cells.
The grid is uniform with the highest resolution in the vicinity of the body and stretched to the external boundaries with a constant ratio (see Table \ref{tab:grid_specs} for details about the grid).
The discretization of the immersed boundary has the same resolution as the background fluid grid.
A convective condition was used at the outlet boundary while freestream conditions were enforced on the three other boundaries.
The job was submitted to a pool of two \texttt{NC24r} nodes, using $12$ MPI processes and $2$ GPU devices per node.
It completed $200,000$ time steps (i.e., $80$ time units of flow simulation with time-step size $\Delta t = 0.0004 c / U_\infty$) in just above $7$ hours (wall-clock time).

\begin{figure}[!h]
    \centering
    \includegraphics[width=8cm]{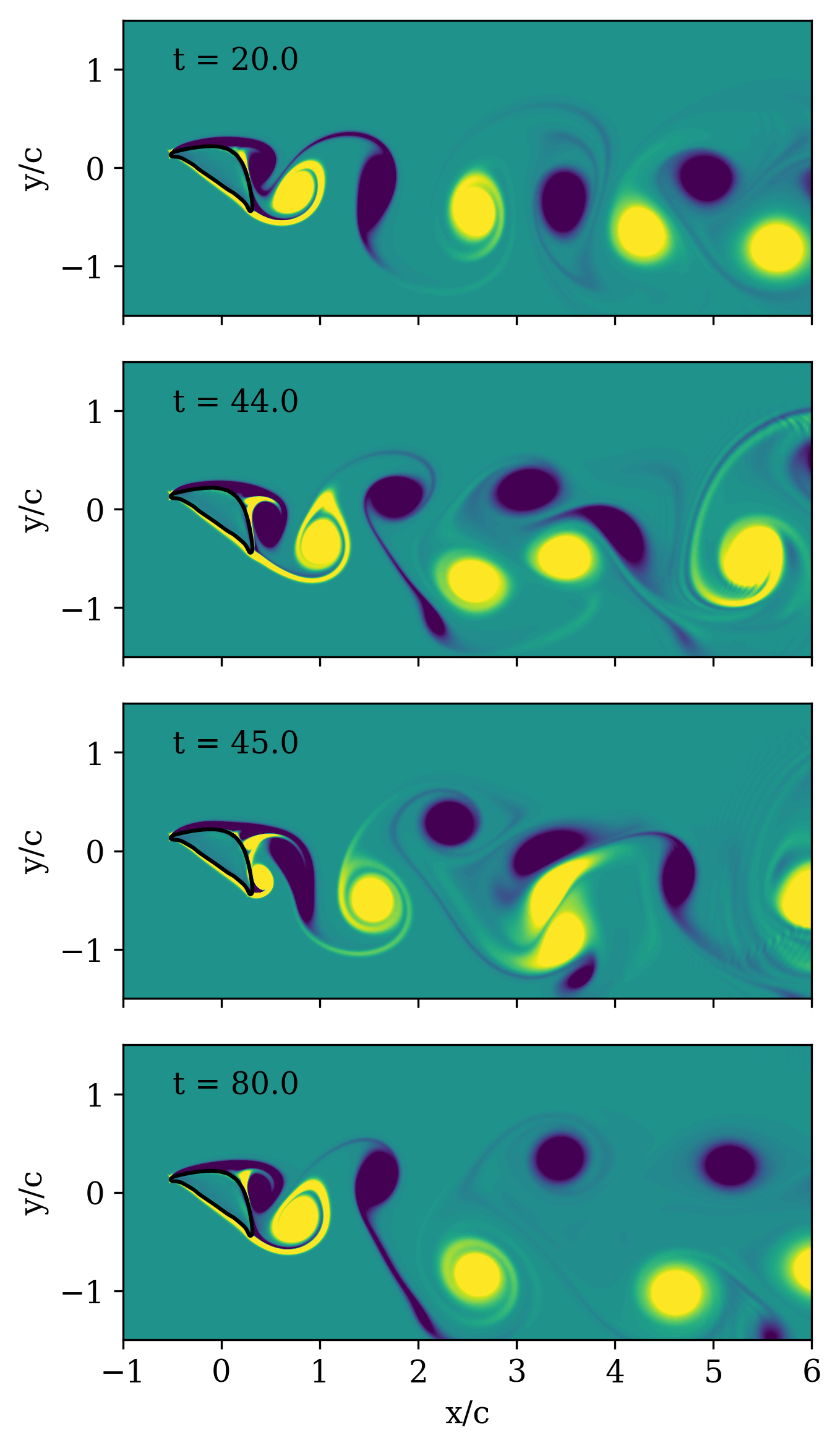}
    \caption{Filled contour of the vorticity field ($-5 \leq w_z c / U_\infty \leq 5$) after $20$, $44$, $45$, and $80$ time units of flow simulation with PetIBM for the snake cross-section at a $35$-degree angle of attack and Reynolds number $2000$. Vortex merging events trigger a change in the wake signature causing the drop in the mean value of the lift coefficient.}
    \label{fig:wz_2d}
\end{figure}

Fig. \ref{fig:force_coefficients} shows the history of the force coefficients on the two-dimensional cross-section.
The lift coefficient only maintains its maximum mean value during the early stage of the simulation (up to $40$ time units).
Between $40$ and $50$ time units, the mean value starts to drop.
The time-averaged force coefficients (between $40$ and $80$ time units) are reported in Table \ref{tab:force_coefficients}.
Fig. \ref{fig:wz_2d} shows snapshots of the vorticity field after $20$, $44$, $45$, and $80$ time units of flow simulation.
After $20$ time units, the vortices shed from the snake section are almost aligned in the near wake (with a slight deflection towards the lower part of the domain).
Snapshots of the vorticity at time units $44$ and $45$ show that the initial alignment is perturbed by vortex-merging events (same-sign vortices merging together to form a stronger one).
Following that, the wake signature is altered for the rest of the simulation: vortices are not aligned anymore, the wake becomes wider (leading to lower aerodynamic forces) with a 1S+1P vortex signature (a single clockwise-rotating vortex on the upper part and a vortex pair on the lower part).

\subsubsection{3D Flow Around a Snake Cylinder}

Although the two-dimensional simulations can give us some insights into to flow dynamics happening in the wake behind the snake, we know that at this Reynolds number ($Re = 2000$), three-dimensional structures will develop in the wake.
We submitted jobs on Azure Batch to perform direct numerical simulation of the three-dimensional flow around a snake model: a cylinder with the same anatomically accurate cross-section.
The computational domain extends in the $z$-direction over a length of $3.2c$ and the grid now contains about $46$ million cells (with a uniform discretization in the $z$-direction and periodic boundary conditions; see Table \ref{tab:grid_specs}).
The job was submitted to a pool of two \texttt{NC24r} nodes using $24$ MPI processes and $4$ GPU devices per node.
The task completed $100,000$ time steps ($100$ time units with a time-step size $\Delta t = 0.01 c / U_\infty$) in about $5$ days and $16$ hours.

\begin{figure}
    \centering
    \includegraphics[width=8.5cm]{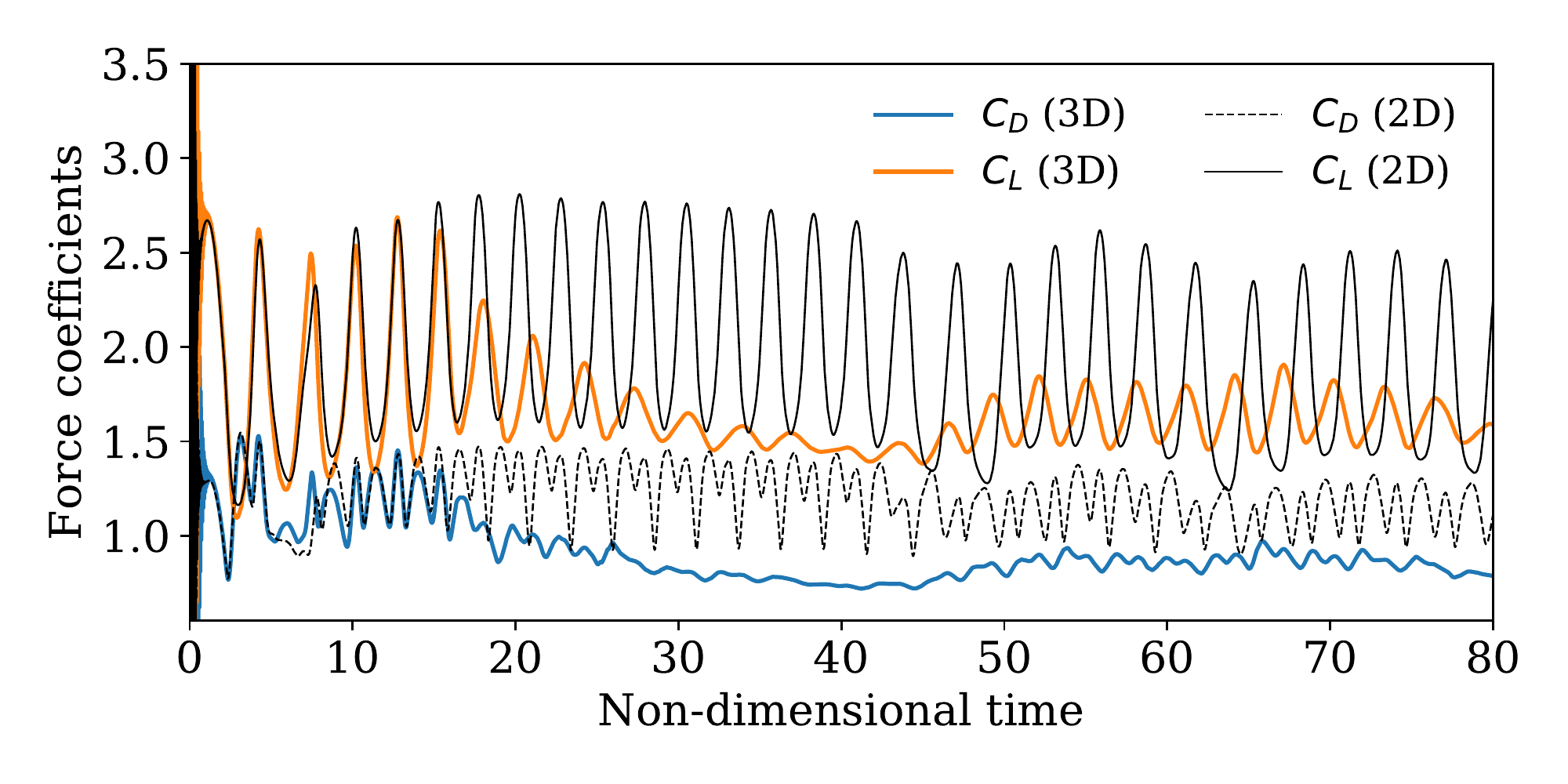}
    \caption{History of the force coefficients obtained with two- and three-dimensional simulations at Reynolds number $2000$ for a snake cross-section with a $35$-degree angle of attack. As expected, the two-dimensional simulations over-estimate the force coefficients of what is essentially a three-dimensional flow problem.}
    \label{fig:force_coefficients}
\end{figure}

Fig. \ref{fig:force_coefficients} compares the history of the force coefficients between the two- and three-dimensional configurations.
The force coefficients resulting from the two-dimensional simulation of the snake are higher than those obtained for the snake cylinder; we computed a relative difference of $+37.9\%$ and $+14.4\%$ for the time-averaged drag and lift coefficient, respectively (see Table \ref{tab:force_coefficients}).
Two-dimensional computational simulations of fundamentally three-dimensional flows lead to incorrect estimation of the force coefficients, as is well known\cite{mittal_balachandar_1995}.
The grid used for the three-dimensional simulation contains about $46$ million cells and is somewhat coarse for direct numerical simulation at this Reynolds number, but acceptable for our exploratory analysis.
A finer grid would be preferable to capture the flow dynamics in detail, but the purpose of this paper is not targeting the physics.
We did however run a simulation on a finer grid (about $233$ million cells) resulting in a relative difference (with respect to the ``coarse''-grid simulation) of $+6.5\%$ and $+3\%$ for the time-averaged drag and lift coefficients, respectively.
We decided to continue this work with the coarser grid due to the large cost difference (see Section \ref{sec:cost}).
Details about this fine-grid simulation are available as supplementary material on the GitHub repository for this study.\footnote{\url{https://nbviewer.jupyter.org/github/barbagroup/cloud-repro/blob/master/misc/independence/grid.ipynb}}

\begin{figure}
    \centering
    \includegraphics[width=8cm]{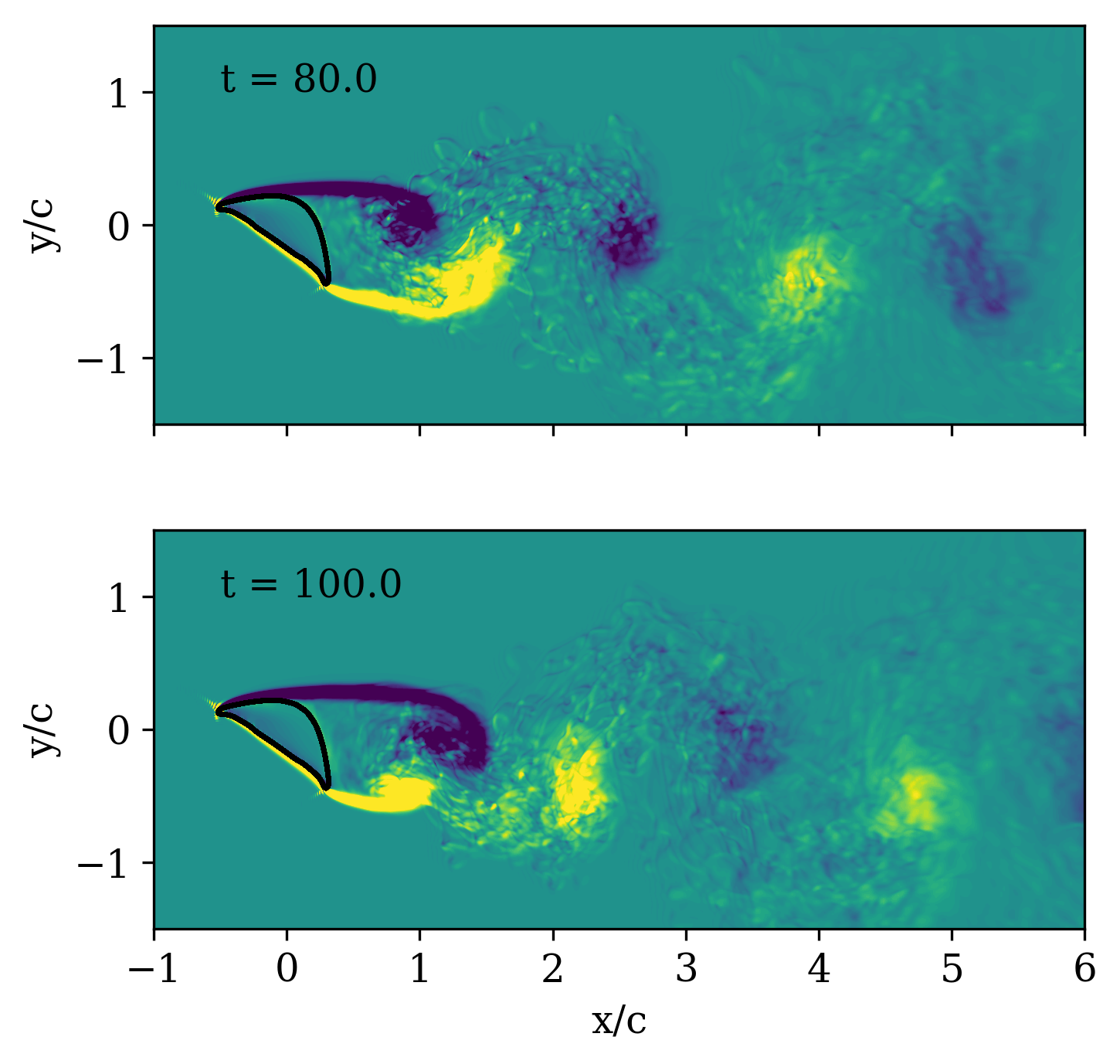}
    \caption{Filled contour of the spanwise-averaged z-component of the vorticity ($-5 \leq w_z c / U_\infty \leq 5$) field after $80$ and $100$ time units of a three-dimensional flow simulation with PetIBM for the snake cylinder with a cross-section at a $35$-degree angle of attack and Reynolds number $2000$.}
    \label{fig:wz_avg_3d}
\end{figure}

\begin{figure}
    \centering
    \includegraphics[width=8cm]{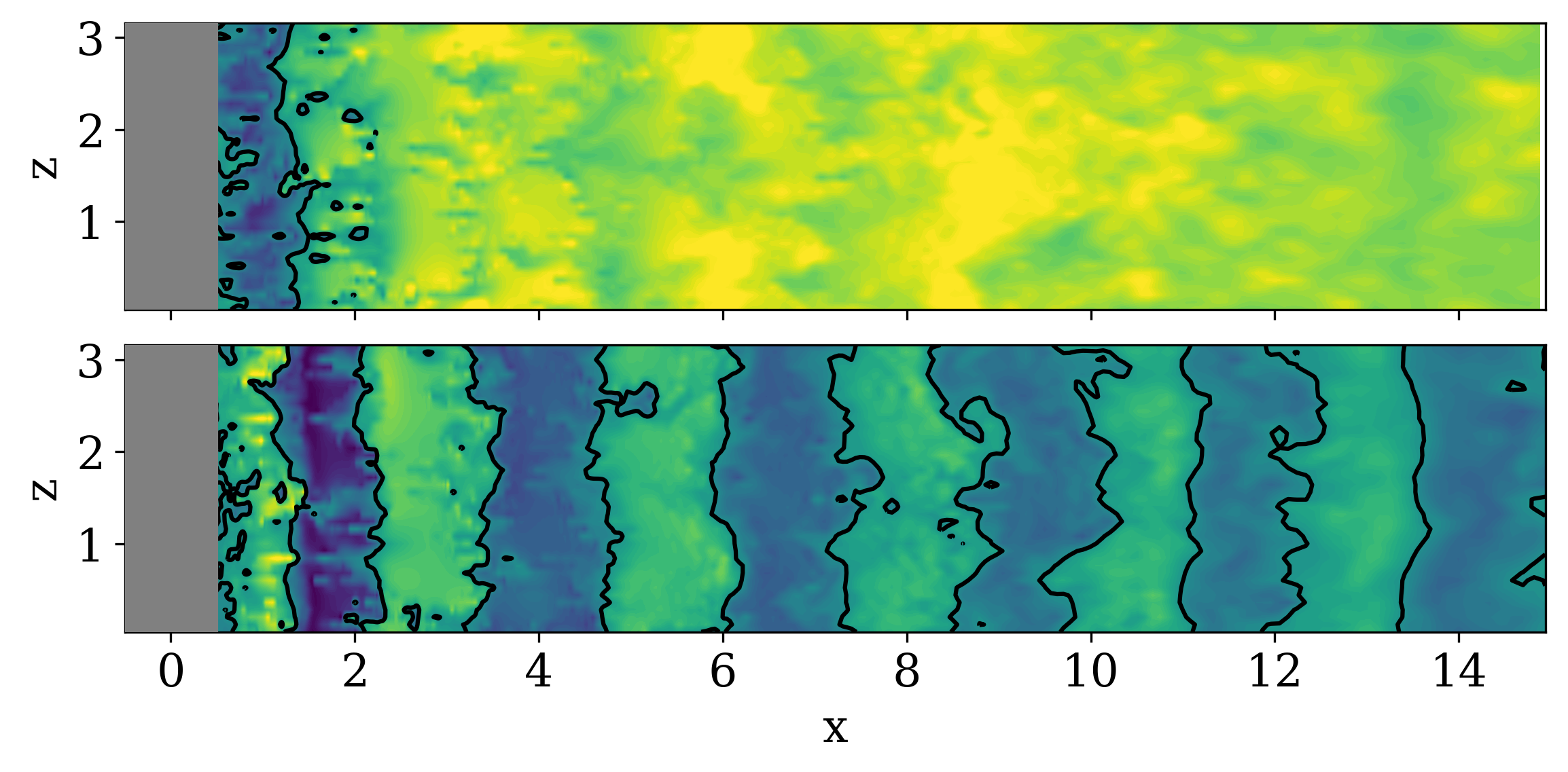}
    \caption{Filled contour of the streamwise velocity (top) and cross-flow velocity (bottom) behind the snake cylinder in the $x/z$ plane at $y/c=-0.2$ in the wake of the snake cylinder with a $35$-degree angle of attack at Reynolds number $2000$ after $100$ time-units of flow simulations. There are 52 contours from $-1.0$ to $1.0$. The grey area shows a projection of the snake cylinder in the $x/z$ plane. The solid black line defines the contour with $u_x = 0$ (top) and $u_y = 0$ (bottom).}
    \label{fig:ux_uy_xz_plane_3d}
\end{figure}

\begin{figure*}[!h]
    \centering
    \includegraphics[width=18cm]{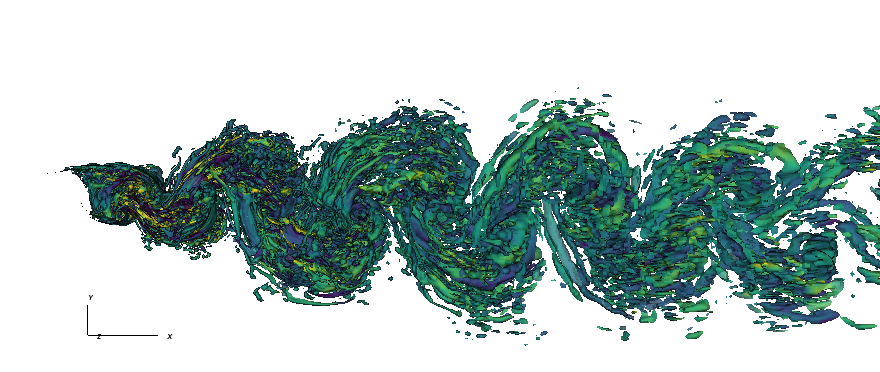}
    \caption{Lateral view of the isosurfaces of the Q-criterion ($Q = 1$) in the wake of the snake cylinder (with a $35$-degree angle of attack) at Reynolds number $2000$. The isosurfaces are colored with the streamwise vorticity ($-5 \leq w_x c / U_\infty \leq 5$). The figure was generated using the visualization software VisIt \cite{childs_et_al_2012}.}
    \label{fig:qcrit_wx_3d}
\end{figure*}

Fig. \ref{fig:wz_avg_3d} shows the instantaneous spanwise vorticity averaged along the $z$-direction after $80$ and $100$ time units.
Compared to the snapshots from the two-dimensional simulation, we observe that free-shear layers roll up into vortices further away from the snake body and that the von K\'{a}rm\'{a}n street exhibits a narrower wake than in the two-dimensional simulations.
We also note the presence of an unsteady recirculation region just behind the snake cylinder, and alternating regions of positive and negative cross-flow velocity showing the presence of von Karman vortices (Fig. \ref{fig:ux_uy_xz_plane_3d}).
Fig. \ref{fig:qcrit_wx_3d} shows a side-view of the isosurfaces of the Q-criterion after $100$ time units and highlights the complexity of the three-dimensional turbulent wake generated by the snake model.

The selection of results presented here corresponds to a typical set of experiments included in a CFD study. 
A comprehensive study might include several similar sets, leading to additional insights about the flow dynamics, and clues to future avenues of research.
We include this selection here to represent a typical workflow, and combine our discussion with a meaningful analysis of the costs associated with running CFD studies in a public cloud.

\begin{table*}
    \renewcommand{\arraystretch}{1.5}
    \caption{Details about the computational grids used for the snake simulations with PetIBM. Distances are expressed in terms of chord-length units.}
    \label{tab:grid_specs}
    \centering
        \begin{tabular}{clllll}
        Case & Domain & Uniform region & Smallest cell-width & Stretching ratio & Size \\
        \hline
        2D & $30 \times 30$ & $\left[ -0.52, 3.48 \right] \times \left[ -2, 2 \right]$ & $0.004 \times 0.004$ & $1.01$ & $1704 \times 1706$ \\
        3D & $30 \times 30 \times 3.2$ & $\left[ -0.52, 3.48 \right] \times \left[ -2, 2 \right] \times \left[ 0, 3.2 \right]$ & $0.008 \times 0.008 \times 0.08$ & $1.01$ & $1071 \times 1072 \times 40$ \\
        \hline
    \end{tabular}
\end{table*}

\begin{table}[!h]
    \renewcommand{\arraystretch}{1.5}
    \caption{Time-averaged force coefficients on the snake model at Reynolds number $2000$ and angle of attack $35^o$ for the two- and three-dimensional configurations. (We average the force coefficients between $40$ and $80$ time units of flow simulation and report the relative difference of the 2D values with respect to the 3D ones.)}
    \label{tab:force_coefficients}
    \centering
    \begin{tabular}{cll}
        Case & $<C_D>$ & $<C_L>$ \\
        \hline
        3D & $0.8390$ & $1.5972$ \\
        2D & $1.1567$ ($+37.9\%$) & $1.8279$ ($+14.4\%$) \\
        \hline
    \end{tabular}
\end{table}

\section{Cost Analysis and User Experience}\label{sec:cost}

For the year $2018$, Microsoft Azure granted our research lab a sponsorship (in terms of cloud credits) to run CFD simulations with our in-house software.
With a ``pay-as-you-go'' subscription, users can immediately see how much it costs to run a scientific application on a public cloud; such information is often hidden to end-users on university-managed clusters.
From May to December, we spent a total of $20,614$ USD to run CFD simulations, including a couple dozens of snake simulations.
(The output of those simulations is now being processed to further analyze the complex flow dynamics generated behind the snake model.)
During that period, we have been charged for data management, networking, storage, bandwidth, and virtual machines (Table \ref{tab:azure_charges}).
More than $99\%$ of the charges incurred were for the usage of virtual machines, mainly instances from the NC-series (Table \ref{tab:nc_series}).
To run PetIBM simulations of the snake model, we used the \texttt{NC24r} virtual machines to get access to NVIDIA K80 GPUs and InfiniBand networking.
For example, the two-dimensional snake run reported in the present article cost $55.4$ USD ($2$ \texttt{NC24r} instances with a hourly price of $3.96$ USD for about $7$ hours).
The three-dimensional computation cost $1077.1$ USD ($2$ \texttt{NC24r} instances for about $136$ hours).
(Note that with 3-year-reserved instances, the two- and three-dimensional snake simulations would have only cost $24.6$ and $478.1$ USD, respectively.)
The three-dimensional simulation on the finer grid ($233$ million cells) ran on $6$ \texttt{NC24r} instances and cost about $7965$ USD to compute $100,000$ time steps (cost not included in \ref{tab:azure_charges}).
This is more than $7$ times the cost we paid for the ``coarse''-grid run!
It would have been too expensive to run a batch of simulations on the finer grid for our exploratory analysis of the flow around the snake cylinder.

\begin{table}
    \renewcommand{\arraystretch}{1.5}
    \caption{Charges incurred for the usage of different services on Microsoft Azure.}
    \label{tab:azure_charges}
    \centering
    \begin{tabular}{cll}
        Service name & Cost (USD) & \% of total cost \\
        \hline
        Bandwidth & $46.85$ & $0.23$ \\
        Data Management & $0.56$ & $0.003$ \\
        Networking & $1.38$ & $0.007$ \\
        Storage & $25.93$ & $0.16$ \\
        Virtual Machines & $20,582.64$& $99.6$ \\
        \hline
        Total & $20,614$ & \\
        \hline
    \end{tabular}
\end{table}

\begin{table*}[!h]
    \renewcommand{\arraystretch}{1.5}
    \caption{NC series on Microsoft Azure. (Prices as of March 24, 2019, for CentOS or Ubuntu Linux Virtual Machines in the East US region.)}
    \label{tab:nc_series}
    \centering
    \begin{tabular}{ccccccccc}
        Instance & cores & \begin{tabular}{@{}c@{}}RAM \\ (GiB) \end{tabular} & \begin{tabular}{@{}c@{}}disk sizes \\ (GiB) \end{tabular} & GPU & \begin{tabular}{@{}c@{}c@{}}pay-as-you-go \\ (dedicated) \\ (\$/hr) \end{tabular} & \begin{tabular}{@{}c@{}c@{}}pay-as-you-go \\ (low-priority) \\ (\$/hr) \end{tabular} & \begin{tabular}{@{}c@{}}1-year reserved \\ (\$/hr) \end{tabular} & \begin{tabular}{@{}c@{}}3-year reserved \\ (\$/hr) \end{tabular} \\
        \hline
        NC6 & 6 & 56 & 340 & 1 x K80 & 0.90 & 0.18 & 0.5733 & 0.3996 \\
        NC12 & 12 & 112 & 680 & 2 x K80 & 1.80 & 0.36 & 1.1466 & 0.7991 \\
        NC24 & 24 & 224 & 1,440 & 4 x K80 & 3.60 & 0.72 & 2.2932 & 1.5981 \\
        NC24r\footnotemark & 24 & 224 & 1,440 & 4 x K80 & 3.96 & 0.792 & 2.5224 & 1.7578 \\
        \hline
    \end{tabular}
\end{table*}
\footnotetext{The NC24r configuration provides a low latency, high throughput network interface optimized for tightly coupled parallel applications.}


Running CFD simulations on Microsoft Azure was a first time for our research lab.
As novices in cloud computing, it took us several months to become familiar with the technical vocabulary and the infrastructure of Microsoft Azure before we could submit our first simulation of the flow around a snake profile.

Command-line utilities such as Azure CLI and Batch Shipyard were of great help to create and manage resources on Microsoft Azure for executing our reproducible research workflow.
(In our lab, we tend to avoid interacting directly with graphical-user interfaces to keep a trace of the commands run and make the workflow more reproducible.)
Azure CLI helped us set up the Azure Batch and Storage accounts as well as moving data between the cloud platform and our local machines.
Thanks to Batch Shipyard, we did not have to dig into the software development kit of Azure Batch to use the service.
Writing YAML configuration files was all we needed to do to create a pool of virtual machines with Batch Shipyard and to submit jobs to it.
With Batch Shipyard, we have painlessly submitted jobs on Azure Batch to run multi-instance tasks in Docker containers, everything from the command-line terminal on a local machine.
Note that Batch Shipyard also supports Singularity containers.
Singularity is an open source container platform designed for HPC workloads.
Singularity containers can be used on traditional HPC clusters such as Colonial One; Docker containers usually cannot.
Indeed, running a Docker container involves running a Docker daemon (a background process) which requires root privileges (that users do not and should not have on production clusters).
The container technology from Singularity was designed from the ground up to prevent escalation in user privileges.

Our in-house CFD software, PetIBM, relies on MPI to run applications on distributed-memory architectures.
The Poisson system is solved on distributed GPU devices using the NVIDIA AmgX library.
Thus, we used Azure instances of the NC-series, which feature a network interface for remote direct memory access (RDMA) connectivity, allowing nodes in the same pool to communicate over InfiniBand network.
The Docker images were specifically built to be able to run MPI applications on RDMA-capable virtual machines (such as the NC24r instance with a CentOS-based 7.3 HPC image) with multiple-instance tasks on Azure Batch service.
As of this writing, only Intel MPI 5.x versions are supported with the Azure Linux RDMA drivers.
Both the Docker image and the configurations files for deploying the cloud resources give an interested reader the opportunity to reproduce our computations in an identical environment.
Even if this opportunity is not accompanied by portability---e.g., one may depend on a different MPI library to use a different system---the human-readable Docker files and configuration files do offer transparency.

Microsoft Azure offers the possibility of taking advantage of surplus capacity with ``low-priority'' virtual machines that are substantially cheaper than ``dedicated'' ones (see Table \ref{tab:nc_series} for pricing options on the NC series with Linux-based virtual machines).
For example, the low-priority \texttt{NC24r} instance costs $0.792$ USD per hour, $5$ times cheaper than its dedicated counterpart.
The reader should keep in mind that low-priority virtual machines may not be available for allocation or may be preempted at any time.
Thus, low-priority instances should be avoided for long-running MPI jobs in pools where inter-node communication is enabled.
We only used dedicated virtual machines to run our CFD simulations.
Moreover, our job tasks used a shared filesystem (GlusterFS on compute nodes) for I/O operations and Batch Shipyard would fail to create the pool if low-priority nodes were requested.

Microsoft Azure implements quotas and limits on resources with Azure Batch service, such as the maximum number of dedicated cores that can be used in a certain region.
To be able to run our simulations, we had to contact Azure Support through Azure Portal to request a quota increase for a given type of instances in a specific region and for a given subscription.
We had to go through this process at least five times during our sponsorship period.
Readers should be aware of these quotas and limits before scaling up workloads on Azure Batch.

\bigskip

\IEEEPARstart{E}{xploring} public cloud services as an alternative (or complementary) solution to university-managed HPC clusters to run CFD simulations with our in-house software, we used Microsoft Azure to generate reproducible computational results.
Our reproducible workflow makes use of Docker containers to faithfully port the runtime environment of our CFD application to Azure,
and the command-line utilities Azure CLI and Batch Shipyard to run multi-instance tasks within Docker containers on the Azure Batch platform.
We ran CFD simulations on instances of the NC-series that feature NVIDIA K80 GPU devices and have access to InfiniBand network.
Latency and bandwidth micro-benchmarks show that our university-managed HPC system and Microsoft Azure can deliver similar performances.
Thanks to a Microsoft Azure sponsorship, we were able to run dozens of CFD simulations of a gliding snake's model.
Our study of the three-dimensional flow around a cylinder with an anatomically accurate cross-section of the flying snake is currently ongoing. 
We plan to analyze the output data from these runs, and publish them in a later paper.

In this work, we show that public cloud resources are today able to deliver similar performances to a university-managed cluster, and thus can be regarded as a suitable solution for research computing.
But in addition to performance, researchers also worry about cost.
We share the actual costs incurred to run two- and three-dimensional CFD simulations  using cloud services with a pay-as-you-go model, and report on reduced costs possible with reserved instances.
Universities (and possibly funding agencies) may obtain even more favorable pricing through bids or medium-term contracts.
Researchers thinking of adding cloud computing to proposal budgets might encounter other barriers, however.
For example, some universities exempt large equipment purchases from indirect (facilities and administrative, F\&A)  costs, while they may add these overhead rates to cloud purchases. 
Until these internal policies are adjusted, cloud computing may not be adopted widely.

In a heightened effort to make our research transparent and reproducible by others, we maintain in a public version-controlled repository all the files necessary to re-run the examples highlighted in this paper.
The repository is found at \url{https://github.com/barbagroup/cloud-repro} and is also permanently archived in Zenodo \cite{cloud_repro_2019}.
Our expanded reproducibility packages contain input files, Batch Shipyard configuration files, command-line instructions to set up and submit jobs to Microsoft Azure, as well as the post-processing scripts to reproduce the figures of this paper.
We included in the Zenodo archive all secondary data required to reproduce the figures without running the simulations again.
Although they are not required to reproduce the computations, the Dockerfiles to build Docker images are also made available. 
We even include a Dockerfile to reproduce the local environment with the visualization tools needed for post-processing.
These files contain details about all the libraries needed to create the computational runtime environment. 
Both the Dockerfiles and the workflow files may not execute as they were designed to do in a period of a few years, due to technology changes.
Obsolescence of digital research objects, including software libraries and computational environments, is an acknowledged source of non-reproducibility, in the long run. 
The recent report of the National Academies highlights this \cite[p.~57]{nasem_2019}. 
Indeed, tools in our workflow that we have no control over (Azure CLI, Batch Shipyard) will likely change, perhaps breaking backward compatibility. 
That is inevitable. 
But when they do, our fully documented workflow (via human-readable configuration files) will continue to offer transparency of what we did and how we did it.

\section*{Acknowledgments}

This work was possible thanks to a sponsorship from the Microsoft Azure for Research program\footnote{Microsoft Azure for Research: \url{https://www.microsoft.com/en-us/research/academic-program/microsoft-azure-for-research/}} and supported by NSF Grant No. CCF-1747669.
We would like to thank Kenji Takeda, Fred Park, and Joshua Poulson of Microsoft for constructive interactions and listening to our feedback about their products. 
We are also immensely grateful to the reviewers, Patrick O'Leary \cite{oleary_2019} and Freddie Witherden \cite{witherden_2019}, for their contributions, which helped make the paper's audience and purpose more clear, in addition to other improvements.
We include our detailed replies to the reviewer suggestions as a markdown file in the GitHub repository for this paper (exported from GitHub issues).

\textbf{Olivier Mesnard} is a doctoral student at the George Washington University. His research interests include computational fluid dynamics and immersed boundary methods with application to animal locomotion. Mesnard has an Engineering degree from Institut Sup{\'e}rieur de M{\'e}canique de Paris and an MS from Universit{\'e} Pierre et Marie Curie. Contact him at mesnardo@gwu.edu.

\textbf{Lorena A. Barba} is a professor of mechanical and aerospace engineering at the George Washington University. Her research interests include computational fluid dynamics, biophysics, and high-performance computing. She is co-Editor of the CiSE Reproducible Research Track, Associate Editor for The ReScience Journal, Associate Editor-in-Chief of the Journal of Open Source Software, and Editor-in-Chief of the Journal of Open Source Education. Barba received a PhD in aeronautics from the California Institute of Technology. Contact her at labarba@gwu.edu.

\bibliographystyle{IEEEtran}

\begin{thebibliography}{10}
\providecommand{\url}[1]{#1}
\csname url@samestyle\endcsname
\providecommand{\newblock}{\relax}
\providecommand{\bibinfo}[2]{#2}
\providecommand{\BIBentrySTDinterwordspacing}{\spaceskip=0pt\relax}
\providecommand{\BIBentryALTinterwordstretchfactor}{4}
\providecommand{\BIBentryALTinterwordspacing}{\spaceskip=\fontdimen2\font plus
\BIBentryALTinterwordstretchfactor\fontdimen3\font minus
  \fontdimen4\font\relax}
\providecommand{\BIBforeignlanguage}[2]{{%
\expandafter\ifx\csname l@#1\endcsname\relax
\typeout{** WARNING: IEEEtran.bst: No hyphenation pattern has been}%
\typeout{** loaded for the language `#1'. Using the pattern for}%
\typeout{** the default language instead.}%
\else
\language=\csname l@#1\endcsname
\fi
#2}}
\providecommand{\BIBdecl}{\relax}
\BIBdecl

\bibitem{donoho_et_al_2009}
D.~L. Donoho, A.~Maleki, I.~U. Rahman, M.~Shahram, and V.~Stodden,
  ``Reproducible research in computational harmonic analysis,'' \emph{Computing
  in Science \& Engineering}, vol.~11, no.~1, 2009.

\bibitem{barba_2018}
L.~A. Barba, ``Terminologies for reproducible research,'' \emph{arXiv preprint
  arXiv:1802.03311}, 2018.

\bibitem{schwab_et_al_2000}
M.~Schwab, N.~Karrenbach, and J.~Claerbout, ``Making scientific computations
  reproducible,'' \emph{Computing in Science \& Engineering}, vol.~2, no.~6,
  pp. 61--67, 2000.

\bibitem{peng_2011}
R.~D. Peng, ``Reproducible research in computational science,'' \emph{Science},
  vol. 334, no. 6060, pp. 1226--1227, 2011.

\bibitem{nasa_oss_2018}
\BIBentryALTinterwordspacing
{National Academies of Sciences, Engineering, and Medicine}, \emph{Open Source
  Software Policy Options for NASA Earth and Space Sciences}.\hskip 1em plus
  0.5em minus 0.4em\relax Washington, DC: The National Academies Press, 2018.
  [Online]. Available: \url{https://doi.org/10.17226/25217}
\BIBentrySTDinterwordspacing

\bibitem{nasem_2019}
\BIBentryALTinterwordspacing
------, \emph{Reproducibility and Replicability in Science}.\hskip 1em plus
  0.5em minus 0.4em\relax Washington, DC: The National Academies Press, 2019.
  [Online]. Available:
  \url{https://www.nap.edu/catalog/25303/reproducibility-and-replicability-in-science}
\BIBentrySTDinterwordspacing

\bibitem{boettiger_2015}
C.~Boettiger, ``An introduction to {Docker} for reproducible research,''
  \emph{ACM SIGOPS Operating Systems Review}, vol.~49, no.~1, pp. 71--79, 2015.

\bibitem{freniere_et_2016}
C.~Freniere, A.~Pathak, M.~Raessi, and G.~Khanna, ``The feasibility of
  {Amazon's} cloud computing platform for parallel, {GPU}-accelerated,
  multiphase-flow simulations,'' \emph{Computing in Science \& Engineering},
  vol.~18, no.~5, p.~68, 2016.

\bibitem{mesnard_barba_2017}
\BIBentryALTinterwordspacing
O.~Mesnard and L.~A. Barba, ``Reproducible and replicable computational fluid
  dynamics: it's harder than you think,'' \emph{Computing in Science \&
  Engineering}, vol.~19, no.~4, p.~44, 2017. [Online]. Available:
  \url{https://doi.org/10.1109/MCSE.2017.3151254}
\BIBentrySTDinterwordspacing

\bibitem{balay_et_al_2018}
\BIBentryALTinterwordspacing
S.~Balay, S.~Abhyankar, M.~F. Adams, J.~Brown, P.~Brune, K.~Buschelman,
  L.~Dalcin, A.~Dener, V.~Eijkhout, W.~D. Gropp, D.~Kaushik, M.~G. Knepley,
  D.~A. May, L.~C. McInnes, R.~T. Mills, T.~Munson, K.~Rupp, P.~Sanan, B.~F.
  Smith, S.~Zampini, H.~Zhang, and H.~Zhang, ``{PETS}c users manual,'' Argonne
  National Laboratory, Tech. Rep. ANL-95/11 - Revision 3.10, 2018. [Online].
  Available: \url{http://www.mcs.anl.gov/petsc}
\BIBentrySTDinterwordspacing

\bibitem{socha_2011}
\BIBentryALTinterwordspacing
J.~J. Socha, ``Gliding flight in chrysopelea: Turning a snake into a wing,''
  \emph{Integrative and Comparative Biology}, vol.~51, no.~6, pp. 969--982,
  2011. [Online]. Available: \url{https://dx.doi.org/10.1093/icb/icr092}
\BIBentrySTDinterwordspacing

\bibitem{chuang_et_al_2018}
\BIBentryALTinterwordspacing
P.-Y. Chuang, O.~Mesnard, A.~Krishnan, and L.~A. Barba, ``{PetIBM}: toolbox and
  applications of the immersed-boundary method on distributed-memory
  architectures,'' \emph{The Journal of Open Source Software}, vol.~3, no.~25,
  p. 558, May 2018. [Online]. Available:
  \url{https://doi.org/10.21105/joss.00558}
\BIBentrySTDinterwordspacing

\bibitem{perot_1993}
J.~B. Perot, ``An analysis of the fractional step method,'' \emph{Journal of
  Computational Physics}, vol. 108, no.~1, pp. 51--58, 1993.

\bibitem{li_et_al_2016}
R.-Y. Li, C.-M. Xie, W.-X. Huang, and C.-X. Xu, ``An efficient immersed
  boundary projection method for flow over complex/moving boundaries,''
  \emph{Computers \& Fluids}, vol. 140, pp. 122--135, 2016.

\bibitem{chuang_barba_2017}
\BIBentryALTinterwordspacing
P.-Y. Chuang and L.~A. Barba, ``{AmgXWrapper}: An interface between {PETSc} and
  the {NVIDIA} {AmgX} library,'' \emph{The Journal of Open Source Software},
  vol.~2, no.~16, Aug. 2017. [Online]. Available:
  \url{https://doi.org/10.21105/joss.00280}
\BIBentrySTDinterwordspacing

\bibitem{mittal_balachandar_1995}
R.~Mittal and S.~Balachandar, ``Effect of three-dimensionality on the lift and
  drag of nominally two-dimensional cylinders,'' \emph{Physics of Fluids},
  vol.~7, no.~8, pp. 1841--1865, 1995.

\bibitem{childs_et_al_2012}
H.~Childs, E.~Brugger, B.~Whitlock, J.~Meredith, S.~Ahern, D.~Pugmire,
  K.~Biagas, M.~Miller, C.~Harrison, G.~H. Weber, H.~Krishnan, T.~Fogal,
  A.~Sanderson, C.~Garth, E.~W. Bethel, D.~Camp, O.~R\"{u}bel, M.~Durant, J.~M.
  Favre, and P.~Navr\'{a}til, ``{VisIt: An End-User Tool For Visualizing and
  Analyzing Very Large Data},'' in \emph{{High Performance
  Visualization--Enabling Extreme-Scale Scientific Insight}}, Oct. 2012, pp.
  357--372.

\bibitem{cloud_repro_2019}
\BIBentryALTinterwordspacing
O.~Mesnard and L.~A. Barba, ``Cloud-repro: Reproducible workflow on a public
  cloud for computational fluid dynamics,'' Zenodo archive, April 2019.
  [Online]. Available: \url{https://doi.org/10.5281/zenodo.2642710}
\BIBentrySTDinterwordspacing

\bibitem{oleary_2019}
\BIBentryALTinterwordspacing
P.~O'Leary, ``Review: "reproducible workflow on a public cloud for
  computational fluid dynamics".'' [Online]. Available:
  \url{https://doi.org/10.22541%2Fau.156415458.85375575}
\BIBentrySTDinterwordspacing

\bibitem{witherden_2019}
\BIBentryALTinterwordspacing
F.~Witherden, ``Review of reproducible workflow on a public cloud for
  computational fluid dynamics,'' Jul 2019. [Online]. Available:
  \url{https://figshare.com/articles/Review_of_Reproducible_Workflow_on_a_Public_Cloud_for_Computational_Fluid_Dynamics/9159740/1}
\BIBentrySTDinterwordspacing

\end{thebibliography}
%

\end{document}